\documentclass[twoside,english,5p,sort&compress,hidelinks]{elsarticle}
\usepackage[T1]{fontenc}
\usepackage[latin9]{inputenc}
\usepackage{babel}
\usepackage{array}
\usepackage{booktabs}
\usepackage{units}
\usepackage{textcomp}
\usepackage{multirow}
\usepackage{amsmath}
\usepackage{graphicx}
\usepackage[unicode=true]
 {hyperref}

\makeatletter

\providecommand{\tabularnewline}{\\}

\journal{Journal of Non-Crystalline Solids. Publisher version at \href{http://dx.doi.org/10.1016/j.jnoncrysol.2019.02.006}{dx.doi.org/10.1016/j.jnoncrysol.2019.02.006}}

\usepackage{lineno}

\usepackage[bracket-numbers = false,
  table-align-uncertainty = false, table-align-exponent=true]{siunitx}
\usepackage[version=3]{mhchem}

\usepackage{hyphenat}

\makeatother

\begin{document}

\begin{frontmatter}{}

\title{Solving the Classical Nucleation Theory with respect to the surface
energy}

\tnotetext[t1]{© 2019. This manuscript version is made available under the CC-BY-NC-ND
4.0 license \href{http://creativecommons.org/licenses/by-nc-nd/4.0/}{http://creativecommons.org/licenses/by-nc-nd/4.0/}}

\author{Daniel R.~Cassar}

\ead{daniel.r.cassar@gmail.com}

\cortext[cor1]{Corresponding author}

\address{Department of Materials Engineering, Federal University of São Carlos,
São Carlos, Brazil}
\begin{abstract}
An essential parameter of the Classical Nucleation Theory (CNT) is
the surface energy between a critical-size nucleus and the ambient
phase, $\sigma$. In condensed matter, this parameter cannot be experimentally
determined independently of CNT. A common practice to obtain $\sigma$
is to assume a model for its temperature-dependence and perform a
regression of the CNT equation against experimental nucleation data.
The drawback of this practice is that assuming the temperature-dependence
of $\sigma$ adds a bias to the analysis. Nonetheless, this practice
is common because an analytical solution of the Classical Nucleation
Theory with respect to $\sigma$ is not possible considering common
expressions of this theory. In this article, a general numerical solution
to this problem using the Lambert $W$ function is proposed, tested,
and compared with typical regression methods. The major advantage
of the proposed method is that there is no need to assume a model
for the temperature-dependence of $\sigma$.
\end{abstract}
\begin{keyword}
Classical Nucleation Theory \sep surface energy \sep Lambert $W$
function
\end{keyword}

\end{frontmatter}{}

\section{Classical Nucleation Theory}

The Classical Nucleation Theory (CNT) formulation shown in Eq.~\eqref{eq:cnt_general}
is based on the work of Gibbs~\citep{Gibbs_1876_equilibrium} with
further modifications from other authors~\citep{Volmer_1926_Nucleation,Farkas_1927_Keimbildungsgeschwindigkeit,Kaischew_1934_Zur,Zeldovich_1942_Contribution,Frenkel_1946_Kinetic}.
In this equation, $J$ is the nucleation rate, $J_{0}$ is a pre-exponential
factor, $D$ is the effective diffusion coefficient related to the
mobility of the units that participate in the nucleation process,
$Z$ is the Zeldovich factor~\citep{Zeldovich_1942_Contribution},
$W^{*}$ is the work of formation of a critical-size nucleus, $k$
is the Boltzmann constant, and $T$ is the absolute temperature.

\begin{equation}
J=J_{0}DZ\exp\left(-\frac{W^{*}}{kT}\right)\label{eq:cnt_general}
\end{equation}

Some expressions exist for the parameters $J_{0}$, $D$, $Z$, and
$W^{*}$~\citep{Schmelzer_2005_Nucleation}. These expressions depend
on specific considerations about the nucleation process. For instance,
if the critical-size nucleus is considered to be spherical and isotropic,
then

\[
W^{*}=\frac{16\pi\sigma^{3}}{3\Delta G_{V}^{2}}\,,
\]
where $\sigma$ is the surface energy between the nucleus and the
ambient phase, and $\Delta G_{V}$ is the change in the Gibbs free
energy (per unit of volume) when the ambient phase transforms into
the phase of the nucleus~\citep{Kelton_1991_Crystal,Cassar_2016_Crystallization}.

Eqs.~\eqref{eq:cnt_visco} and \eqref{eq:cnt_tau} show two expressions
for Eq.~\eqref{eq:cnt_general} found in the literature~\citep{Kelton_1991_Crystal,Schmelzer_2005_Nucleation,Fokin_2006_Homogeneous,Gupta_2016_Role,Gupta_2016_variation},
considering a plethora of assumptions discussed therein.

\begin{equation}
J=\frac{\sqrt{kT\sigma}}{\eta\lambda^{5}}\exp\left(-\frac{16\pi\sigma^{3}}{3kT\Delta G_{V}^{2}}\right)\label{eq:cnt_visco}
\end{equation}

\begin{equation}
J=\frac{16\sqrt{kT\sigma^{3}}}{3\tau\Delta G_{V}^{2}\lambda^{6}}\exp\left(-\frac{16\pi\sigma^{3}}{3kT\Delta G_{V}^{2}}\right)\label{eq:cnt_tau}
\end{equation}

In the previous expressions, $\eta$ is the shear viscosity of the
ambient phase, $\lambda$ is the jump distance for diffusion estimated
by the size of the units that diffuse during the nucleation process,
and $\tau$ is the time-lag of the nucleation process.

Eqs.~\eqref{eq:cnt_visco} and \eqref{eq:cnt_tau} cannot be solved
analytically for $\sigma$. The objective of this short communication
is to show how to solve this kind of problem numerically using the
Lambert $W$ function~\citep{Corless_1996_LambertW}.

The domain of Eqs.~\eqref{eq:cnt_visco} and \eqref{eq:cnt_tau}
is dependent on the assumptions used to obtain them; here they will
be used only for experimental nucleation rate data above the reported
temperature of maximum nucleation. Fortunately, the method proposed
here is general and can also be used for other expressions of $J$.

\section{Lambert $W$ function}

The Lambert $W$ function (also known as the omega function) is defined
as the inverse function of $f(z)=z\exp(z)$, with $z$ being any complex
number, thus

\[
z=W_{n}(z\exp(z))\,.
\]

\begin{figure}
\begin{centering}
\includegraphics[width=0.75\columnwidth]{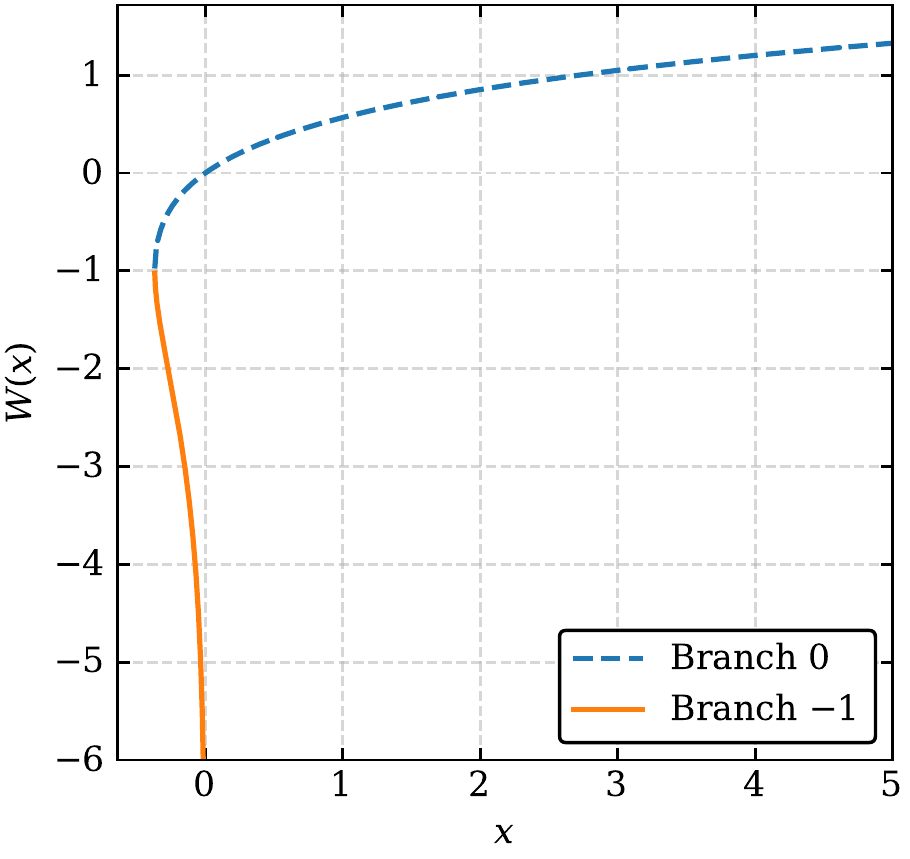}
\par\end{centering}
\caption{Real values of the $W$ functions computed in the branches \num{0}
and \num{-1}.\label{fig:plotW}}
\end{figure}

The $W$ is a multivalued function with $n$ branches, and one must
select an integer $n$ to solve it. Only $n=\num{0}$ and \num{-1}
produce real results, the only ones of interest to the problem here,
considering there is no physical meaning for complex values of $\sigma$.
Figure \ref{fig:plotW} shows the real values of $W$ for these two
branches.

Eqs.~\eqref{eq:cnt_visco} and \eqref{eq:cnt_tau} can be written
in the form

\begin{equation}
a\sigma^{p}\exp(b\sigma^{q})=1\,,\label{eq:sigma_eq}
\end{equation}
where $a$, $p$, $b$, and $q$ are non-zero parameters that depend
on the nucleation expression in consideration (see Table~\ref{tab:parametros}).
A solution of Eq.~\eqref{eq:sigma_eq} for $\sigma$ is

\begin{equation}
\sigma=\left[\frac{p}{bq}W_{n}\left(\frac{bq}{p}a^{-\frac{q}{p}}\right)\right]^{\frac{1}{q}}\,.\label{eq:sigma_solve}
\end{equation}

\begin{table}
\caption{Parameters $a$, $p$, $b$, and $q$ for Eqs.~\eqref{eq:cnt_visco}
and \eqref{eq:cnt_tau}. The parameters $b$ and $q$ are the same
for both equations.\label{tab:parametros}}

\centering{}%
\begin{tabular}{lcccc}
\toprule 
 & $a$ & $p$ & $b$ & $q$\tabularnewline
\midrule
\midrule 
Eq.~\eqref{eq:cnt_visco} & $\sqrt{kT}\left[J\eta\lambda^{5}\right]^{-1}$ & $\nicefrac{1}{2}$ & \multirow{2}{*}{{\large{}$-\frac{16\pi}{3\Delta G_{V}^{2}kT}$}} & \multirow{2}{*}{$3$}\tabularnewline
Eq.~\eqref{eq:cnt_tau} & $16\sqrt{kT}\left[3J\Delta G_{V}^{2}\tau\lambda^{6}\right]^{-1}$ & $\nicefrac{3}{2}$ &  & \tabularnewline
\bottomrule
\end{tabular}
\end{table}

With the solution shown in \eqref{eq:sigma_solve}, it is possible
to solve expressions \eqref{eq:cnt_visco} and \eqref{eq:cnt_tau}
for $\sigma$. In fact, any equation that can be written in the form
of \eqref{eq:sigma_eq} can be solved for $\sigma$. This solution
may also be used, for example, when dealing with crystal growth mediated
by secondary surface nucleation (see Ref.~\citep{Cassar_2017_Elemental}).
The fact that a solution is possible does not imply that it is non-complex.
In some cases, a real solution may be impossible.

Considering that $\sigma$ is expected to increase monotonically with
respect to the temperature in the range where nucleation data are
available~\citep{Greer_1991_Nucleation}, the only physically valid
branch of $W$ in this scenario is branch \num{-1}.

Several free and commercial numerical programs have an implementation
of the $W$ function. Scipy~\citep{Jones_2001_SciPy}, for instance,
is a free and open-source module for the Python programming language
that has this function built in with the name \texttt{lambertW}, available
in the \texttt{scipy.special} submodule. For an iterative approach
that can be performed using a spreadsheet program, see Ref.~\citep{Allan_2012_Inverting}.

\section{Test on experimental data}

Table~\ref{tab:dados_coletados} shows crystal nucleation data measured
in a supercooled \ce{Li2B4O7} liquid (lithium tetraborate, also known
as lithium diborate), as well as other properties of interest~\citep{Cassar_2014_Crystal}.
With these data and the solution shown in Eq.~\eqref{eq:sigma_solve},
the value of $\sigma$ was computed considering the two expressions
for $J$ that are common in the literature (Eqs.~\eqref{eq:cnt_visco}
and \eqref{eq:cnt_tau}). The results are shown in Table~\ref{tab:dados_calculos}.
The computed $\sigma$ values are within the expected range for this
parameter (\num{0.1} to \SI{0.25}{J.m^{-2}}) ~\citep{Deubener_1998_Crystalliquid,Fokin_2000_Crystal,Fokin_2000_Method,Fokin_2006_Homogeneous,Schmelzer_2019_Curvature}.

\begin{table}
\caption{Crystal nucleation data and properties of interest of \ce{Li2B4O7}~\citep{Cassar_2014_Crystal}.
Uncertainty in $J$, $\tau$, and $\eta$ is one standard deviation.
Uncertainty in $J$ and $\tau$ is the standard deviation obtained
in the parameters of the non-linear regression of the Collins\textendash Kashchiev
equation~\citep{Collins_1955_Time,Kashchiev_1969_Solution} (see~\citep{Cassar_2014_Crystal}
for more details). Uncertainty in $\eta$ is the confidence band of
the shear viscosity regression.\label{tab:dados_coletados}}

\centering{}%
\begin{tabular}{ccccc}
\toprule 
$T$ & $J\times\num{e-7}$ & $\tau$ & $\Delta G_{V}\times\num{e-8}$ & $\eta\times\num{e-9}$\tabularnewline
{[}\si{K}{]} & {[}\si{m^{-3}.s^{-1}}{]} & {[}\si{s}{]} & {[}\si{J.m^{-3}}{]} & {[}\si{Pa.s}{]}\tabularnewline
\midrule
\midrule 
773 & 90(1) & 1323(8) & \num{-4.20} & 54(2)\tabularnewline
783 & 64(1) & 877(9) & \num{-4.15} & 7.0(2)\tabularnewline
793 & 6.9(1) & 140(10) & \num{-4.10} & 1.10(4)\tabularnewline
\bottomrule
\end{tabular}
\end{table}

\begin{table}
\caption{Values of $\sigma$ computed using data from Table~\ref{tab:dados_coletados},
considering $\lambda=$~\SI{4.86e-10}{m} and uncertainty of \SI{2}{K}
in $T$. $T$ in \si{K} and $\sigma$ in \si{J.m^{-2}}. Values obtained
using branch \num{-1} of the Lambert $W$ function. \label{tab:dados_calculos}}

\centering{}%
\begin{tabular}{ccc}
\toprule 
 & Eq.~\eqref{eq:cnt_visco} & Eq.~\eqref{eq:cnt_tau}\tabularnewline
$T$ & $\sigma$ & $\sigma$\tabularnewline
\midrule
\midrule 
773 & 0.1621(2) & 0.1608(1)\tabularnewline
783 & 0.1649(2) & 0.1614(2)\tabularnewline
793 & 0.1697(2) & 0.1664(2)\tabularnewline
\bottomrule
\end{tabular}
\end{table}

\section{Comparison with regression methods}

Another approach to obtain $\sigma$ from experimental $J(T)$ data
is via regression. A method popularized by James \citep{James_1985_Kinetics}
consists of linearizing Eq.~\eqref{eq:cnt_visco} by plotting $\ln\left(J\eta T^{-1/2}\right)$
versus $\left[T\Delta G_{V}^{2}\right]^{-1}$, with the underlying
assumption that $\sigma$ and $\lambda$ are temperature-independent.
With a linear regression of the data in the linearized plot, $\sigma$
can be obtained from the slope of the regression, which is equal to
$-16\pi\sigma^{3}$$\left[3k\right]^{-1}$, and $\lambda$ can be
obtained from the intercept of the regression, which is equal to $\ln\left(\sqrt{kT\sigma}\lambda^{-5}\right)$.

Values of $\sigma=$~\SI{0.32(1)}{J.m^{-2}} and $\lambda=$~\SI{1(5)e-32}{m}
were obtained using this method with the data shown in Table~\ref{tab:dados_coletados}.
The value of $\sigma$ is two times higher than the value obtained
using the Lambert $W$ function. The value of $\lambda$ is 22 orders
of magnitude smaller than what is expected, which is a discrepancy
that is well documented in the literature \citep{Kelton_1991_Crystal,Fokin_2006_Homogeneous}.
Rarely discussed, however, is the uncertainty in $\lambda$ when using
this method, which is massive for the dataset investigated here.

As a consistency test, Table~\ref{tab:method_test} shows the computed
value of $J$ when putting the values of $\sigma$ and $\lambda$
obtained using this method back into Eq.~\eqref{eq:cnt_visco}. The
computed value is shown in the column ``Method 1''. The difference
between the reported and the computed value is up to \SI{42}{\percent}
for the temperature of \SI{783}{K}.

\begin{table}
\caption{Reported values of $J$ and calculated values of $J$ using different
methods. See the text for a description of each method. $T$ in \si{K}
and $J$ in \si{m^{-3}.s^{-1}}.\label{tab:method_test}}

\centering{}%
\begin{tabular}{ccccc}
\toprule 
 & \multicolumn{4}{c}{$J\times\num{e-7}$}\tabularnewline
\cmidrule{2-5} \cmidrule{3-5} \cmidrule{4-5} \cmidrule{5-5} 
$T$ & Reported & Method 1 & Method 2 & Method 3\tabularnewline
\midrule
\midrule 
773 & 90(1) & 110 & 90 & 90\tabularnewline
783 & 64(1) & 45 & 64 & 64\tabularnewline
793 & 6.9(1) & 8.1 & 31 & 6.9\tabularnewline
\bottomrule
\end{tabular}
\end{table}

A similar approach using Eq.~\eqref{eq:cnt_tau} instead of Eq.~\eqref{eq:cnt_visco}
was proposed by Weinberg and Zanotto \citep{Weinberg_1989_Re}, which
yields similar results (not shown here). One advantage of these two
regression approaches (advanced by James, Weinberg, and Zanotto) is
that no assumption must be made on the value of $\lambda$. However,
the resulting uncertainty in $\lambda$ is enough to discard this
apparent advantage in this case. Moreover, assuming $\sigma$ to be
temperature-independent may be a reasonable approximation only for
sufficiently small temperature ranges.

Another regression approach is to assume some model for the temperature-dependence
of $\sigma$ and use it to perform a non-linear regression of CNT.
The expression $\sigma=\num{2.8e-4}T-\num{0.054}$ is obtained if
we consider Eq.~\eqref{eq:cnt_visco}, $\lambda=$~\SI{4.86e-10}{m},
and assume that $\sigma$ has a linear temperature\hyp{}dependence.
The computed values of $J$ using this method are shown in the column
``Method 2'' in Table~\ref{tab:method_test}. The difference between
the reported and the computed value is \SI{78}{\percent} for the
temperature of \SI{793}{K}.

Finally, Table~\ref{tab:method_test} also shows the calculated values
of $J$ using the method proposed in this article, column ``Method
3''. It is no surprise that these values are equal to the experimental
values because no assumption concerning the temperature-dependence
of $\sigma$ is necessary to use this new method; thus it gives the
exact value of $\sigma$ to bring theory and experiment to an agreement,
point by point. The hope is that with fewer assumptions we can reach
new insights into this problem.

\section{Conclusion}

An analytical solution for the surface energy may not be possible
depending on the form of Classical Nucleation Theory that is used.
Here I demonstrate a numerical solution (using the Lambert $W$ function)
to solve the Classical Nucleation Theory with respect to the surface
energy. The advantage of this numerical approach, in contrast with
previous methods, is that no assumption on the temperature-dependence
of the surface energy is necessary.

\section*{Acknowledgements}

Daniel R.~Cassar gratefully acknowledges the São Paulo Research Foundation
for the financial support (FAPESP grant number 2017/12491-0) and Carolina
B.~Zanelli for the English revision.

\bibliographystyle{elsarticle-num}
\bibliography{bibliography}

\end{document}